\documentclass[prl,twocolumn,superscriptaddress,floatfix,preprintnumbers]{revtex4-1}
\usepackage{graphics,amssymb,amsmath,epsfig,color}
\usepackage{graphicx}

\begin{document}

\title{The Role of SrTiO$_3$ Phonon Penetrating into thin FeSe Films in the Enhancement of Superconductivity}

\author{Shuyuan Zhang}
\author{Jiaqi Guan}
\author{Xun Jia}
\author{Bing Liu}
\author{Weihua Wang}
\affiliation{Beijing National Laboratory for Condensed Matter Physics and Institute of Physics, Chinese Academy of Sciences, Beijing 100190, China}
\author{Fangsen Li}
\affiliation{State Key Laboratory of Low-Dimensional Quantum Physics, Department of Physics, Tsinghua University, Beijing 100084, China}
\author{Lili Wang}
\email[]{liliwang@mail.tsinghua.edu.cn}
\affiliation{State Key Laboratory of Low-Dimensional Quantum Physics, Department of Physics, Tsinghua University, Beijing 100084, China}
\affiliation{Collaborative Innovation Center of Quantum Matter, Beijing 100871, China}
\author{Xucun Ma}
\author{Qikun Xue}
\affiliation{State Key Laboratory of Low-Dimensional Quantum Physics, Department of Physics, Tsinghua University, Beijing 100084, China}
\affiliation{Collaborative Innovation Center of Quantum Matter, Beijing 100871, China}
\author{Jiandi Zhang}
\author{E. W. Plummer}
\affiliation{Department of Physics and Astronomy, Louisiana State University, Baton Rouge, LA 70808, USA}
\author{Xuetao Zhu}
\email[]{xtzhu@iphy.ac.cn}
\affiliation{Beijing National Laboratory for Condensed Matter Physics and Institute of Physics, Chinese Academy of Sciences, Beijing 100190, China}
\author{Jiandong Guo}
\email[]{jdguo@iphy.ac.cn}
\affiliation{Beijing National Laboratory for Condensed Matter Physics and Institute of Physics, Chinese Academy of Sciences, Beijing 100190, China}
\affiliation{Collaborative Innovation Center of Quantum Matter, Beijing 100871, China}

\date{\today}


\begin{abstract}

The significant role of interfacial coupling on the
superconductivity enhancement in FeSe films on SrTiO$_3$ has been
widely recognized. But the explicit origination of this coupling is
yet to be identified. Here by surface phonon measurements using high
resolution electron energy loss spectroscopy, we found electric
field generated by Fuchs-Kliewer (F-K) phonon modes of SrTiO$_3$ can
penetrate into FeSe films and strongly interact with electrons
therein. The mode-specific electron-phonon coupling (EPC) constant
for the $\sim$92 meV F-K phonon is $ \sim 0.25$ in the single-layer
FeSe on SrTiO$_3$. With increasing FeSe thickness, the penetrating
field intensity decays exponentially, which matches well the
observed exponential decay of the superconducting gap. It is
unambiguously shown that the SrTiO$_3$ F-K phonon penetrating into
FeSe is essential in the interfacial superconductivity enhancement.

\end{abstract}
\maketitle

The recent discovery of high-temperature superconductivity in 1 unit
cell (u.c.) FeSe films on SrTiO$_3$ (STO) substrate
\cite{WangQingyan2012-1ucFeSe,ZhangWenhao2014-CPL-1ucFeSe} has
attracted a lot of attention, since the superconducting transition
temperature (T$_C$) is significantly raised to $\sim 60-75$ K
\cite{ZhangWenhao2014-PhysRevB-1ucFeSe,liuDefa2012FeSe,LeeJJ2014FeSereplica,TanShiyong2013FeSe},
even over 100 K \cite{GeJianfeng2015FeSe}, from 8 K for the bulk
FeSe \cite{hsu2008FeSe}. Various experiments have been performed to
uncover the veiled mechanism of the T$_C$ enhancement in this
system. Although no explicit conclusion has been reached, two
factors are widely believed to be essential -- electron doping to
FeSe films and the coupling at interface between FeSe and STO.

The significance of electron doping in the interfacial
superconductivity enhancement has been evidenced by the high
resolution angle resolved photoemission spectroscopy (ARPES)
measurements. Extensive annealing drives electrons transferred from
STO substrate to the 1 u.c. FeSe thus increases the electron density
in the film, and broadens the superconducting gap accordingly
\cite{liuDefa2012FeSe,HeShaolong2013FeSe,HeJunfeng2014FeSe}. In fact
the superconducting 1 u.c. FeSe/STO has similar Fermi surface as
other two typical electron doped iron-based superconductors,
\emph{e.g.}, A$_x$Fe$_2$Se$_2$ (A=K,Cs) \cite{ZhangYi2011KFeSe} and
(Li,Fe)OHFeSe \cite{ZhaoLin2015LiFeOHFeSe}. The electron density in
thick FeSe films can also be tuned by alkali-metal doping
\cite{Seo2015KdopedFeSe,CHPwen2016NC,Miyata2015KdopedFeSe,SongCanli2016KdopedFeSe}
or by ion-liquid gate tuning
\cite{Shiogai2015gateFeSe,Lei2015gateFeSe,hanzawa2015gateFeSe}. Even
though the electron density can be raised up to the value as high as
that in the 1 u. c. FeSe/STO with T$_C$ of $60 - 75$ K, the
superconductivity in thick FeSe film is always weaker (with the
highest reported T$_C$ of $\sim$48 K \cite{Lei2015gateFeSe}) than
that in the 1 u.c. FeSe/STO. It is indicated that, besides electron
doping, there must be other factor(s) that is responsible for the
superconductivity enhancement in FeSe/STO.

The importance of the interfacial coupling has been directly
evidenced by the substrate selection behavior of the
superconductivity enhancement. On oxide substrates such as SrTiO$_3$
or BiTiO$_3$, T$_C$ is strongly enhanced no matter what crystal
orientation, crystal symmetry or lattice constant the substrates
have
\cite{PengRui2014PRLFeSe,PengRui2014NCFeSe,Zhang2015FeSeSTO110,ZhouGuanyu2015FeSeSTO110};
while when thin FeSe films are grown on graphene/SiC(0001)
\cite{Song2011science,SongCanli2016KdopedFeSe,Zhang2016Nanoletter},
even though similar electron density is accomplished by K-doping,
the maximum superconducting gap is always smaller than that of 1
u.c. FeSe/STO. Moreover, a ferroelectric transition of STO was
observed by Raman spectroscopy at $\sim$50 K, quite close to the
superducting T$_C$ of the 1 u.c. FeSe/STO, implying the possible
correlation between the substrate lattice and the superconductivity
enhancement at the interface \cite{CuiYT2015prl}. Recent high
resolution ARPES measurements show that each energy band of the 1
u.c. FeSe is replicated at $\sim 100$ meV higher binding energy,
which can be attributed to the interaction with the STO optical
phonon around 80 meV \cite{LeeJJ2014FeSereplica}. Theoretical
analyses further suggested that this strong interaction between the
high energy STO phonon and the electrons in FeSe is responsible for
the T$_C$ enhancement \cite{LeeJJ2014FeSereplica}.

All these findings seem to correlate the interfacial coupling with
some unique properties of the oxide substrates, especially the
optical phonon mode. By measuring the properties directly related to
the electronic state, such as the electronic structure or the
electron's life time, the electron-phonon coupling (EPC) constant
can be roughly estimated \cite{LeeJJ2014FeSereplica,Zhao2016}. But
the involved phonon mode cannot be identified directly, which limits
our understanding of the underlying physics. In this letter, we
report on surface phonon measurements for thin FeSe films grown on
STO, obtained by high resolution electron energy loss spectroscopy
(HREELS). From the phonon perspective, our results clearly reveal
that electric field generated by the Fuchs-Kliewer (F-K)
\cite{FK1965} surface phonon modes of STO can penetrate into thin
FeSe films and strongly interact with the electrons in the FeSe
layer with a mode-specific coupling constant value
$\lambda_{\alpha}\sim0.25$ for the $\sim$92 meV branch. The
incomplete screening of the electric field associated with the F-K
modes is the key to induce the interfacial coupling between FeSe and
STO, and further result in the superconductivity enhancement.

Thin FeSe films with different thickness (1 u.c., 2 u.c., 3 u.c. and
10 u.c.) were grown on STO (001) surface by molecular beam epitaxy
following the procedures reported in Ref.
\cite{WangQingyan2012-1ucFeSe,ZhangWenhao2014-CPL-1ucFeSe}. For
convenience, these samples will be referred to as 1uc-FeSe/STO,
2uc-FeSe/STO, etc., respectively. All samples were characterized by
scanning tunneling microscopy to confirm the high growth quality.
And the superconducting property of 1uc-FeSe/STO was verified by
scanning tunneling spectroscopy and ARPES. The determined
superconducting gap ($\Delta$) is around 20 meV and T$_C$ is in the
range of 60-70 K. The details about the sample preparation and
characterization are described in the Supplementary Material
\cite{Supplementary}.

\begin{figure}
\begin{center}
\includegraphics[width=0.5\textwidth]{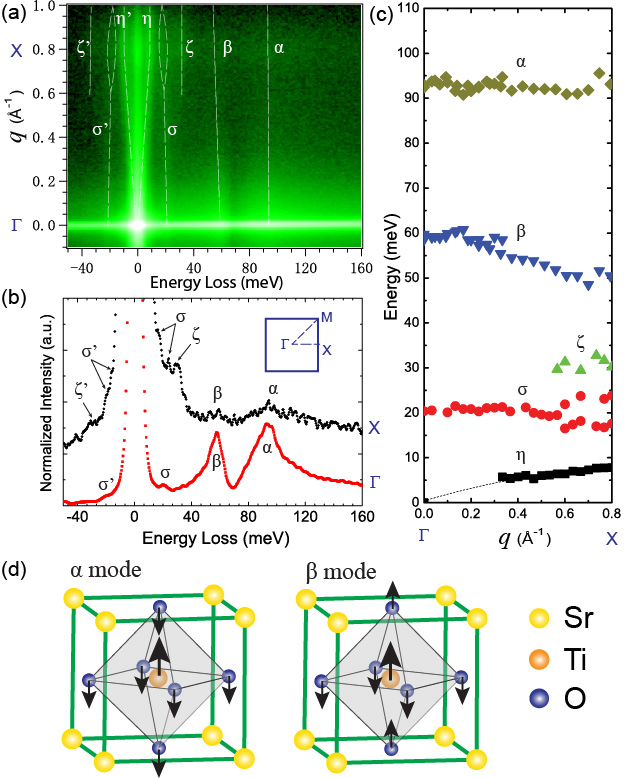}
\caption{\label{fig1}{\small HREELS results of the 1uc-FeSe/STO
sample at room temperature. (a) 2D energy-momentum mapping along the
$\Gamma$-X direction. Five different phonon modes with positive
energy loss are labeled by $\alpha$, $\beta$, $\sigma$, $\zeta$, and
$\eta$, respectively. Dashed lines are provided to guide the eye.
The corresponding negative energy loss features correspond to their
anti-Stokes peaks, which are labeled by $\sigma'$, $\eta'$, etc. (b)
EDCs at $\Gamma$ point and X point with the inset shows the
Brillouin zone. (c) Phonon dispersions from the results of panel(a).
(d) Ionic vibrations of the $\alpha$ and $\beta$ modes
\cite{Vogt1988}.}}
\end{center}
\end{figure}

As a surface sensitive technique, HREELS is an ideal tool to explore
the substrate effects on epitaxial thin films. HREELS measurements,
carried out by the recently developed 2D-HREELS system
\cite{ZhuXuetao2015HREELS}, were performed on all samples in the
temperature range from 35 K to 300 K. Fig. \ref{fig1} shows the
HREELS results of 1uc-FeSe/STO at room temperature. Five different
energy loss features are observed. The assignments
\cite{Supplementary} of these features are shown in Table.
\ref{tb1}.

\begin{table}
 \centering
\caption{The assignments of the energy loss features
\label{tb1}}\vspace{0.02in}
 \begin{tabular}{c|c|c}\hline\hline
     Feature                   &  Energy ($\hbar\omega$)                    & Assignment      \\ \hline
     $\eta$                    & $\sim 0-7$ meV                             &  acoustic phonon of FeSe  \\
     $\sigma$                  & $\sim 20$ meV                              &  optical phonon of FeSe \\
     $\zeta$                   & $\sim 32$ meV                              &  optical phonon of FeSe  \\
     $\beta$                   & $\sim 60$ meV                              &  F-K surface phonon of STO  \\
     $\alpha$                  & $\sim 92$ meV                              &  F-K surface phonon of STO  \\
 \hline\hline
 \end{tabular}
 \end{table}

As exhibited in the energy distribution curves (EDCs) of Fig.
\ref{fig1} (b), the most prominent features are the $\alpha$ and
$\beta$ modes corresponding to the F-K surface phonons of STO
\cite{Baden1981}. Normally the phonons of substrate covered by a
metal film should not have been detected by HREELS. But these F-K
phonon modes are always accompanied with large dipole oscillations
with electric field extending out of the STO surface \cite{FK1965},
as if the F-K modes could penetrate into the epitaxial metal film.
In HREELS measurement, the incident electrons are so sensitive to
the dipole field \cite{ibach1982eels} that the detection of the
penetrating electric field from the substrate can be realized. The
penetration makes it possible that the electrons in FeSe films
interact with the STO phonons. Moreover, as demonstrated in Fig.
\ref{fig1} (c), the energy of the $\alpha$ mode is dispersionless,
\emph{i.e.} almost a constant of $\sim92$ meV within the
experimental resolution of about 3 meV at different momentum values,
which corroborates the ARPES observation \cite{LeeJJ2014FeSereplica}
that the replica band is separated from the main band by $\sim 100$
meV all through the Brillouin Zone. Our analysis reveals that this
$\alpha$ mode induces rather strong EPC, and thus in the following
we will focus on the $\alpha$ phonon branch as a representative. The
contribution of the $\beta$ mode, with relatively weak EPC, is
described in the Supplementary Material \cite{Supplementary}.

\begin{figure}
\begin{center}
\includegraphics[width=0.5\textwidth]{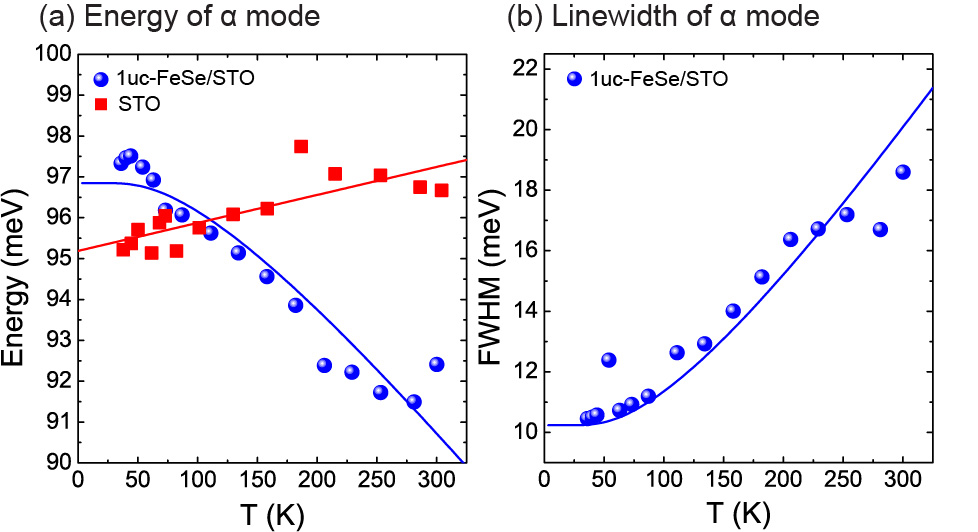}
\caption{\label{fig2} {\small The temperature dependence of the (a)
energy and (b) Full width at half maximum (FWHM) of the $\alpha$
phonon branch (blue: 1uc-FeSe/STO; red: bare STO). The dots are
experimental data obtained by fitting the EDCs from the HREELS
spectra at the $\Gamma$ point with Gaussian peaks. The red solid
line is a linear fitting for bare STO, and the blue solid lines for
1uc-FeSe/STO represent the results of the phonon-phonon decay
fitting described in the text.}}
\end{center}
\end{figure}

The HREELS measurements on 1uc FeSe/STO and bare STO are carried out
at various temperatures and the temperature dependence of the energy
and line width of the $\alpha$ phonon branch are shown in Fig.
\ref{fig2}. The phonon energy of bare STO exhibits a very small
increase with increasing temperature. In particular, when we take
into account an instrument resolution of about 3 meV, it is
negligible. This is consistent with the results in Refs.
\cite{Servoin1980} and \cite{Perry1967STOphonon}. The line width of
bare STO is not shown here because it is overlapped with some other
collective modes at low temperature and can not give credible
results without critical analysis. This will be investigated in
detail elsewhere. In contrast, the phonon energy as well as the line
width of 1uc-FeSe/STO show a very strong temperature dependence.
Obviously the growth of FeSe on STO changes the energy and its
temperature-dependent behavior of the F-K phonon mode drastically.
Since F-K phonon modes are strongly related to the dielectric
response of STO, the observed T-dependence of the F-K mode reflects
the coupling between the F-K phonons and the electrons at the
interface.

The energy for a specific phonon branch $\nu$ at temperature $T$ and
momentum ${\bf q}$ can be written as a complex form
\cite{Grimvall1981}
\begin{equation}\label{eq1}
\widetilde{\omega}({\bf
q},\nu,T)=\texttt{Re}\left[\widetilde{\omega}({\bf
q},\nu,T)\right]-i\texttt{Im}\left[\widetilde{\omega}({\bf
q},\nu,T)\right],
\end{equation}
with the energy as the real part
\begin{equation}\label{eq2}
\texttt{Re}\left[\widetilde{\omega}({\bf
q},\nu,T)\right]=\omega_{0}({\bf q},\nu)+\Delta\omega_{V}({\bf
q},\nu;T)+\Delta\omega_{pp}({\bf q},\nu;T),
\end{equation}
and the line width (half width at half maximum) as the imaginary
part
\begin{equation}\label{eq3}
\texttt{Im}\left[\widetilde{\omega}({\bf
q},\nu,T)\right]=\gamma_{ep}({\bf q},\nu)+\gamma_{pp}(\bf{q},\nu;T),
\end{equation}
where $\omega_{0}(\bf{q},\nu)$ is the harmonic phonon energy at
$T=0$ K including the $T$-independent EPC contribution,
$\Delta\omega_{V}(\bf{q},\nu;T)$ is the energy shift due to the
change in the volume, and $\Delta\omega_{pp}(\bf{q},\nu;T)$ is the
energy shift due to the anharmonic phonon-phonon interactions. The
imaginary term $-i\texttt{Im}\left[\widetilde{\omega}({\bf
q},\nu,T)\right]$ describes the damping of the phonons, with
contributions from both EPC $\gamma_{ep}(\bf{q},\nu)$ and
anharmonicity $\gamma_{pp}(\bf{q},\nu;T)$. Considering the
anharmonic interaction usually lower the phonon energy at an
elevated temperature, the slight temperature-dependence of the
$\alpha$ phonon branch of bare STO should be mainly due to the
volume change. However, when covered by FeSe, the significant energy
decrease with increasing temperature of the $\alpha$ phonon branch
should be attributed to strong anharmonic phonon-phonon interaction.

This anharmonic phonon-phonon interaction can be easily perceived in
Table. \ref{tb1} from the point of view of the energy conservation.
One can immediately tell that the $\alpha$ mode decays into $\zeta$
and $\beta$ modes should be the most favorable decay channel. This
conjecture is verified by simulating the anharmonic interaction via
three different phonon decay models \cite{Supplementary}. Since the
$\alpha$ mode is dispersionless (shown in Fig.\ref{fig1}), we set
the energy to be $\bf{q}$-independent in our models. It turns out
the best fitting results are indeed from a model that the $\alpha$
mode with energy $\hbar \omega_0=92$ meV decays into another two
optical phonon modes ($\zeta$ and $\beta$) with lower energies
$\hbar \omega_1=60$ meV and $\hbar \omega_2=32$ meV, where the
restriction of energy conservation $\hbar \omega_0 = \hbar \omega_1
+ \hbar \omega_2$ is satisfied. In this model the
temperature-dependent phonon energy $\hbar\omega(T)$ and line width
$\gamma(T)$ are expressed as \cite{Menendez1984}:
\begin{equation}\label{eq4}
\hbar\omega(T)=\hbar\omega_{a}+\hbar\omega_{b}(1+\frac{1}{e^{\hbar
\omega_1/k_BT}-1}+\frac{1}{e^{\hbar \omega_2/k_BT}-1}),
\end{equation}
and
\begin{equation}\label{eq5}
\gamma(T)=\gamma_{ep}(1+\frac{1}{e^{\hbar
\omega_1/k_BT}-1}+\frac{1}{e^{\hbar \omega_2/k_BT}-1}),
\end{equation}
where $\hbar\omega_{a}$, $\hbar\omega_{b}$, and $\gamma_{ep}$ are
fitting parameters. The fitting results are plotted as blue solid
lines in Fig. \ref{fig2}. This model explicitly clarifies the big
difference of the anharmonic feature between STO and 1uc-FeSe/STO.
The 32 meV $\zeta$ mode of FeSe, collaborating with the 60 meV
$\beta$ mode of STO, coincidentally renders an appropriate phonon
decay channel for the 92 meV $\alpha$ mode. However, if there is no
FeSe film, this decay channel is absent in STO substrate due to the
restriction of energy conservation. Hence this anharmonic
interaction characterizes part of the interfacial coupling. Its role
in the superconductivity enhancement is not clear yet. Further
studies are needed to figure out the explicit role of the
anharmonicity.

The other part of the interfacial coupling, the penetrating STO F-K
phonons interacting with the electrons in FeSe, is directly relevant
to the superconductivity enhancement. The strength of this
interaction can be characterized by a mode-specific EPC constant
$\lambda({\bf q},\nu)$ for a single phonon mode of wave vector
$\bf{q}$ and branch $\nu$, which is related to the EPC-induced line
width $\gamma_{ep}({\bf q},\nu)$ by Allen's formula \cite{Allen1974,
Allen1980}
\begin{equation}\label{eq6}
\lambda({\bf q},\nu)=\frac{2\gamma_{ep}({\bf q},\nu)}{\pi
N(E_F)\hbar^2\omega^2({\bf q},\nu)},
\end{equation}
where $\omega({\bf q},\nu)$ is the phonon frequency and $N(E_F)$ is
the density of states for both spin in each unit cell at the Fermi
energy.

To obtain the EPC constant from Allen's formula, the EPC-induced
line width $\gamma_{ep}({\bf q},\nu)$ should be a prerequisite
quantity. However, it is challenging to directly obtain the pure EPC
induced phonon line widths from experiment, because the measured
phonon line widths also contain additional contributions from
anharmornic phonon-phonon interactions as shown in Eq.\ref{eq3}.
Only when the temperature-dependent anharmonic phonon-phonon
interaction is correctly deducted, can the EPC-induced line width
$\gamma_{ep}$ be determined and $\lambda$ be calculated. This can be
accomplished from measurements of the temperature dependence of the
phonon dispersions \cite{Grimvall1981} (see Fig. \ref{fig2}). We
extract $\gamma_{ep}$ for $\alpha$ mode through above anharmonic
phonon decay model, as shown in Eq. \ref{eq4} and Eq. \ref{eq5},
which gives $\gamma_{ep}(\alpha)=\gamma(T=0)\cong5.1$ meV, since
there is no phonon-phonon interaction at $T=0$ and $\gamma(T=0)$ is
the line width with EPC only. Independently, by analyzing the
temperature-dependence of the phonon energy, we have also tried
another approach using Kramers-Kronig relation to obtain
$\gamma_{ep}(\alpha)$ \cite{Supplementary}. In this method,
$\gamma_{pp}(\nu;T)$ in Eq. \ref{eq3} can be estimated through the
Kramers-Kronig transformation of $\Delta\omega_{pp}(\nu;T)$ in Eq.
\ref{eq2}, which can be determined from a fitting of experimental
data. Hence by subtracting the anharmonic contribution
$\gamma_{pp}(\alpha;T)$ from measured line width $\gamma(\alpha;T)$,
we obtained $\gamma_{ep}(\alpha)\cong4.5$ meV \cite{Supplementary}.

On the other hand, a simple tight-binding model is applied to fit
the electron band structure near the Fermi Surface to calculate the
density of states $N(E_F)=1.4\times10^{-3} \ ({\textup{meV}})^{-1}$
\cite{Supplementary}. Thus the EPC constant of the $\alpha$ mode
could be obtained from Eq.\ref{eq6}, $\lambda_{\alpha}\sim0.25$.
This coupling constant is significant for such a high energy phonon
mode, which plays a major role in the superconductivity enhancement
\cite{WangYan2016}. The same analysis is performed to the $\beta$
mode and gives $\lambda_{\beta}\sim0.1$. These EPC constants are
mainly attributed to the interactions with electrons in FeSe films,
since the electron density in STO is much lower than FeSe after
electron transfer \cite{TanShiyong2013FeSe}. The sum of the two EPC
constants is $\lambda_{\alpha+\beta}\sim0.35$, which dominantly
accounts for the calculated total EPC constant about 0.4 from STO
substrate \cite{ZhouYuanjun2016}. The higher energy $\alpha$ mode
has a stronger interaction than the $\beta$ mode, as evidenced from
previous ARPES result that the 100 meV replica band is much clearer
than the 60 meV replica band \cite{LeeJJ2014FeSereplica}. This can
also be understood by the fact that the $\alpha$ mode generates
stronger dipole field than the $\beta$ mode. As shown in Fig.
\ref{fig1} (d), all the six oxygen ions vibrate in the opposite
direction with the titanium ions in the $\alpha$ mode, while the two
apical oxgyen ions vibrate in the same direction with the titanium
ions in the $\beta$ mode \cite{Vogt1988}.

\begin{figure}
\begin{center}
\includegraphics[width=0.5\textwidth]{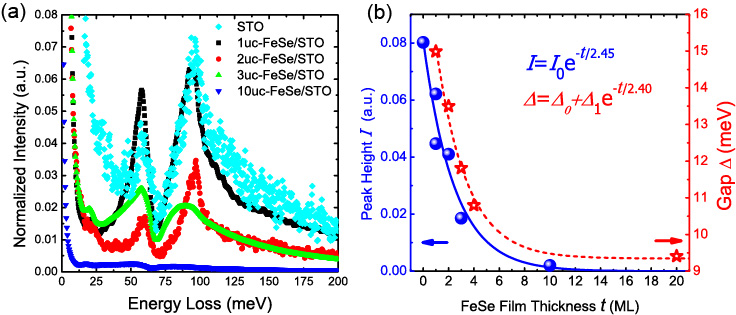}
\caption{\label{fig3} {\small (a) EDCs at the $\Gamma$ point for
samples with different FeSe thickness. (b) Plot and exponential
fitting of the peak height of the $\alpha$ mode as a function of the
FeSe thickness (blue). Plot and exponential fitting of the
superconducting gap size for the K-doped FeSe/STO as a function of
the FeSe thickness (red), with data extracted from Ref.
\cite{Zhang2016Nanoletter}.}}
\end{center}
\end{figure}

Above results give direct evidence that the electric field generated
by the F-K phonon modes of STO can penetrate the FeSe films and
strongly interact with the electrons in the FeSe layer. But how far
can this electric field penetrate? To answer this question we
performed HREELS measurements on samples with different FeSe
thickness (2uc, 3uc and 10uc) to study the spacial spreading
properties of the F-K modes. In Fig.\ref{fig3} (a) we plot the
normalized EDCs at $\Gamma$ point for the samples with different
FeSe thicknesses, which apparently demonstrates the decay of the
$\alpha$ and $\beta$ modes with increasing FeSe thickness. An
exponential fit to the normalized peak height of the $\alpha$ mode
(blue dots in Fig.\ref{fig3} (b)) gives the decay length of 2.5
u.c., which can be regarded as a characteristic length of the F-K
mode penetration in FeSe films. A recent study of the
thickness-dependent superconducting gaps of K-doped FeSe films
presents very similar behavior, \emph{i.e.}, the gap size decreases
exponentially with increasing FeSe thickness characterized by a
decay length of 2.40 u.c. (red stars in Fig.\ref{fig3} (b))
\cite{Zhang2016Nanoletter}. Considering the similar carriers density
at the interfacial layer \cite{Ye2015}, the striking coincidence of
the two characteristic lengthes suggests that the $\alpha$ phonon of
STO is closely related to the superconductivity enhancement at
FeSe/STO interface. The electrons in FeSe spontaneously screen the
F-K electric field and lead to the decay of the field intensity,
which weakens the supconductivity enhancement. This is why the
superconducting gap can only be enhanced within 3 u.c. FeSe films
\cite{TangChenjia2015_1-2ucFeSe, TangChenjia2016FeSe}.

The screening length to an external electric field by a metal is
inversely proportional to the electron density of the metal
\cite{Theophilou1972}. Raising the electron-doping level of the FeSe
film, although helpful to increase the superfluid density, will
inhibit the penetration length of the dipole field from STO.
Therefore there is always a trade-off between the electron density
and the field penetration. This is possibly the cause of gap
decrease when extra electrons are doped in 1uc-FeSe/STO
\cite{TangChenjia2015_1-2ucFeSe}. As a result, increasing the doping
level in FeSe may not be able to further enhance the
superconductivity if the dipole field from the substrate is not
penetrating effectively into the FeSe films. Searching substrates
with F-K phonons that generate strong diploe field is a feasible
route.

In conclusion, our HREELS study provides the microscopic
understanding on the superconductivity enhancement in the FeSe/STO
system. First, we determine the origination of the interfacial
coupling. It is the F-K phonon modes in STO, generating long range
dipole field, that strongly interact with the electrons in the FeSe
layer. The mode-specific EPC constant is $\sim0.25$ for the $\sim$92
meV phonon in 1uc FeSe/STO. This interaction is closely related to
the enhancement of superconductivity. Second, we determine the
characteristic penetration depth of the STO F-K phonon into FeSe
film. The decay of the penetrating phonon results in the thickness
limit of superconductivity enhancement in FeSe/STO. Therefore ionic
crystals with high energy F-K phonon modes that generate strong
dipole field, for instance, oxides with high valence metal ions and
long oxygen-metal bonds, should be good candidates as the
substrates.

X. Zhu acknowledges helpful discussions with Prof. M. El-Batanouny,
and thanks Dr. Yan Wang for pointing out the initial error in
calculating the EPC constant. This work is supported by NSFC (Grant
No. 11225422 and No. 11304367), MOSTC (Grant No. 2012CB921700 and
No. 2015CB921000), the Strategic Priority Research Program (B) of
CAS (Grant No. XDB07010100), and the External Cooperation Program of
BIC, CAS (Grant No. 112111KYSB20130007).


\clearpage \widetext

\begin{centering}
\textbf{\large Supplemental Materials}
\end{centering}

\setcounter{equation}{0} \setcounter{figure}{0}
\setcounter{table}{0} \setcounter{page}{1} \makeatletter
\renewcommand{\theequation}{S\arabic{equation}}
\renewcommand{\thefigure}{S\arabic{figure}}
\renewcommand{\bibnumfmt}[1]{[S#1]}
\renewcommand{\citenumfont}[1]{S#1}

\section{Sample Preparation and Characterization}

\begin{figure}[h!]
\setlength{\abovecaptionskip}{-0.5cm}
\begin{center}
\includegraphics[width=0.6\textwidth]{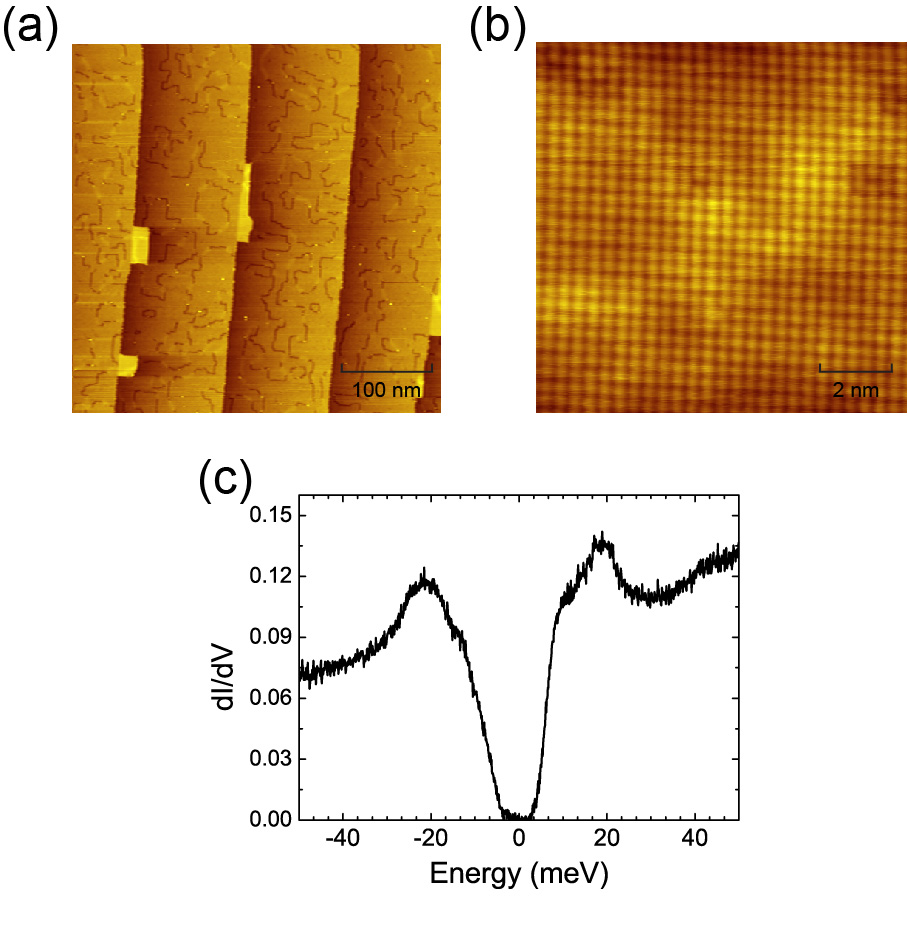}
\caption{\label{figS1}{\small (a) STM topography of 1uc-FeSe/STO
sample (Image size: $400\times400$ nm, sample bias $V_{b}=5.0$ V,
tunneling current $I_{t}=50$ pA). (b) Atomically resolved STM
topography of 1uc-FeSe/STO sample (Image size: $10\times10$ nm,
sample bias $V_{b}=0.4$ V, tunneling current $I_{t}=100$ pA). (c)
dI/dV spectrum of 1uc-FeSe/STO. All Spectra were taken at 4.2 K.}}
\end{center}
\end{figure}

Nb-doped STO(001) substrate was first degassed overnight at 600
$^\circ$C in ultra high vacuum (UHV), and then annealed at 950
$^\circ$C for 40 minutes. FeSe films were grown by co-depositing
high-purity Fe (99.99\%) and Se (99.99+\%) with a flux ratio of
$\sim 1:20$ onto the STO substrate held at 400 $^\circ$C. The
as-prepared samples were post-annealed at 470 $^\circ$C for 5 hours
in UHV to make the first layer FeSe superconducting. The \emph{in
situ} scanning tunneling microscopy (STM) measurements were
performed to confirm the sample quality. Fig. \ref{figS1} (a) gives
the topographic image which shows large surface steps ($\sim100$ nm)
of 1uc-FeSe/STO. The dark strips correspond to domain boundaries in
FeSe films due to the strain between STO substrate and FeSe single
layer. The lattice constant of the FeSe layer is 3.8 \r A as shown
in Fig. \ref{figS1} (b). The scanning tunneling spectroscopy (STS)
measurement in Fig. \ref{figS1} (c) indicates the superconducting
gap ($\Delta$) of annealed 1uc-FeSe/STO is $\sim 20$ meV. The sample
was protected by a thick amorphous Se capping layer deposited at 110
K and transformed to the 2D-HREELS system.

\begin{figure}[h!]
\begin{center}
\includegraphics[width=0.7\textwidth]{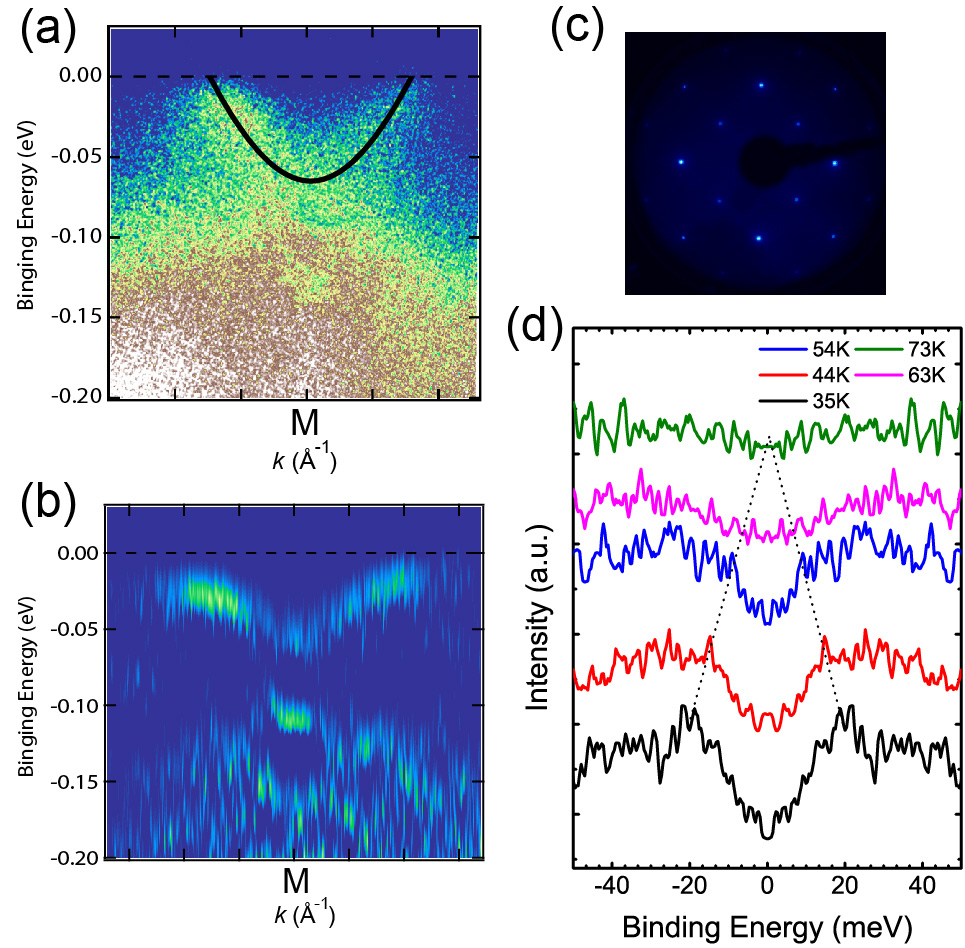}
\caption{\label{figS2} {\small (a) ARPES spectrum of 1uc-FeSe/STO
sample taken at 35 K. $k$ refers to the momentum along the
$\Gamma-M$ direction and centered at $M$. The black line represents
the band fitting by a tight binding model. (b) Second derivative
spectrum for (a). (c) LEED pattern of 1uc-FeSe/STO sample taken at
35 K with 80 eV electron beam energy. (d) Plot of the evolution of
the symmetrized EDCs at $k_F$ as a function of temperature, which
shows the gap closes between 63 K and 73 K.}}
\end{center}
\end{figure}

In the 2D-HREELS system, the sample was annealed at 450 $^\circ$C to
remove the capping layer. ARPES was performed to detect the
electronic structure (Fig. \ref{figS2} (a)) and supercongucting gap
(Fig. \ref{figS2} (d)) to confirm the existence of superconducting
state in the sample. ARPES spectra were scanned along the $\Gamma$
to $M$ direction, corresponding to the horizontal direction in the
LEED pattern in Fig. \ref{figS2} (c). A parabolic electron band can
be clearly observed in the second derivative spectrum (Fig.
\ref{figS2} (b)). Temperature-dependent symmetrized EDCs at $k_{F}$
reveal the superconducting gap at 35 K is $\sim 20$ meV and closes
between 63 K and 73 K (As shown in Fig. \ref{figS2}(d)).

\section{HREELS Experiments and Phonon Mode Assignment}

The HREELS experiments were performed on bare STO and on thin FeSe
films grown on STO with different thickness (1uc, 2uc, 3uc and 10
uc). The STO(001) sample was annealed at 900 $^\circ$C for 1 hour
before the HREELS measurements. Most of the HREELS measurements were
performed with an incident beam energy of 50 eV unless stated
otherwise.

\begin{figure}[h!]
\begin{center}
\includegraphics[width=0.9\textwidth]{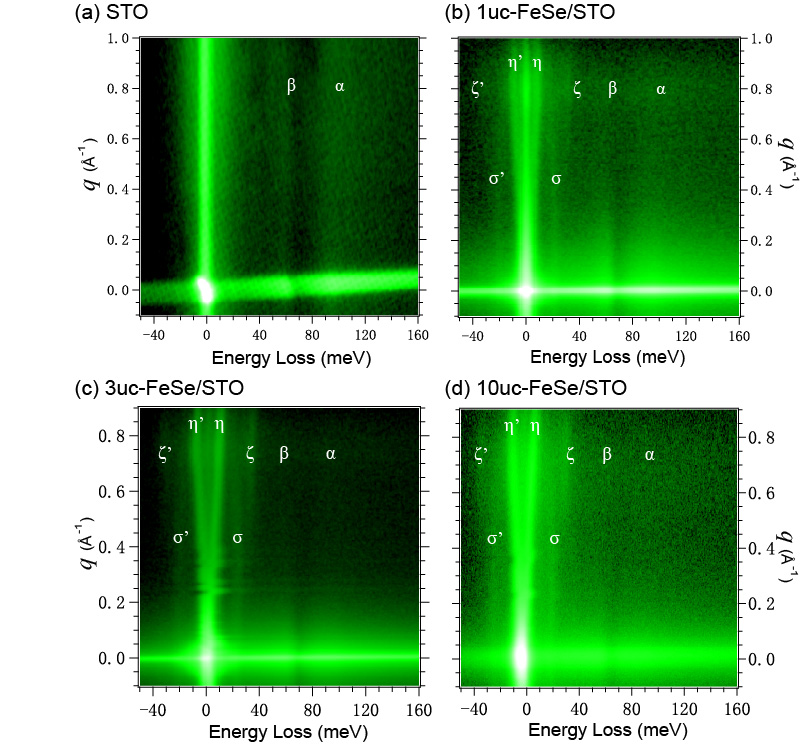}
\caption{\label{figS6} {\small 2D HREELS energy-momentum mapping
along the $\Gamma$-X direction for (a) STO(001), (b) 1uc-FeSe/STO,
(c) 3uc-FeSe/STO and (d) 10uc-FeSe/STO. The incident electron energy
is 120 eV for (a) and 50 eV for (b)-(c)}. Greek letters are used to
label the energy loss features. The positive energy loss features
are labeled by $\alpha$, $\beta$, $\sigma$, $\zeta$, and $\eta$,
respectively. The corresponding negative energy loss features
correspond to their anti-Stokes peaks, which are labeled by
$\sigma'$, $\zeta'$, etc.}
\end{center}
\end{figure}

Fig. \ref{figS6} (a) shows the 2D energy-momentum mapping of the
bare STO(001) sample. There are only two energy loss features at 60
($\beta$ mode) and 97 meV ($\alpha$ mode), corresponding to the F-K
surface phonon modes of STO \cite{FK1965-s}, which is consistent
with previous HREELS measurements
\cite{CONARD1993STOphonon-s,Baden1981-s}. However, for the FeSe/STO
samples, 5 phonon modes are observed. Both theoretical
\cite{Subedi2008FeSephonon-s} and experimental
\cite{Phelan2009FeSephonon-s,Ksenofontov2010FeSephonon-s} studies of
FeSe have shown that the energy of all FeSe phonon modes should be
smaller than 40 meV. Therefore the $\alpha$ (92 meV) and $\beta$ (60
meV) modes observed from FeSe/STO samples should be the F-K modes
from STO substrate. $\eta$, $\sigma$ and $\zeta$ modes, which are
absent in bare STO sample (Fig. \ref{figS6} (a)), are phonon modes
of FeSe lattice. This assignment can also be verified by comparing
the HREELS results of FeSe films with different thickness. Fig.
\ref{figS6} (b), (c) and (d) show the 2D-HREELS mapping of
1uc-FeSe/STO, 3uc-FeSe/STO and 10uc-FeSe/STO, respectively. Clearly
the F-K modes of STO ($\alpha$ and $\beta$) gradually become
invisible with increasing FeSe thickness. In contrast, the $\eta$,
$\gamma$ and $\zeta$ modes become clearer. Actually this is one of
the key reasons why we assign $\eta$, $\gamma$ and $\zeta$ to be the
phonon modes from FeSe, rather than from STO. $\beta$ and $\alpha$
phonon modes observed in FeSe/STO reveals that the electric field
generated by F-K phonon of STO can penetrate into the FeSe films and
decay exponentially with thickness increasing (Fig. 3 in the main
paper).

The detailed assignment of the FeSe phonon modes are based on the
selection rule of HREELS \cite{deJuan2015225-s} and previous Raman
experiments \cite{Okazaki2011FeSeRaman-s,Kumar2010FeSeRaman-s}.
$\eta$ mode is an acoustic phonon branch, $\sigma$ mode is $A_{1g}$
optical phonon, and $\zeta$ should be $E_{g}$ or $A_{2u}$ optical
phonon in FeSe.

\section{Calculation of the EPC constant}

\subsection{Tight binding model and Density of State}

The Density of state (DOS) can be calculated through the electron
band dispersion by a simple tight binding model:
\begin{equation}
\epsilon_{e,h}(k)=-2t_{e,h}\cos(ka)-\mu_{e,h}
\end{equation}
with t$_e$ = 125 meV, t$_h$ = -30 meV, $\mu_e$ = -185 meV and
$\mu_h$ = 175 meV \cite{LeeJJ2014FeSereplica-s}. The fitting of the
band was shown in Fig. \ref{figS2} (a).

The area in k-space for 2D is expressed as:
\begin{equation}
\Omega(E)=\pi k^{2}.
\end{equation}
Therefore, the density of state (DOS) for both spin direction is:
\begin{equation}
N(E)
=\frac{2}{\left(\frac{2\pi}{a}\right)^{2}}\frac{d\Omega(E)}{dE}=\frac{a^{2}}{2\pi}\frac{k}{\frac{d\epsilon_{e}}{dk}}=\frac{a}{2\pi
t_{e}}\frac{k}{\sin(ka)}
\end{equation}

For 1uc-FeSe/STO, a $\sim 3.8$ $\textup{\AA}$, and Fermi wave number
$k_{F}=0.19$ $\textup{\AA}^{-1}$. Therefore,
$N(E_{F})=1.4\times10^{-3}$ (meV)$^{-1}$.

\subsection{Extracting EPC-induced phonon line widths from experimental results: anharmonic phonon decay}

The energy for a specific phonon branch $\nu$ at temperature $T$ and
momentum ${\bf q}$ can be written as a complex form
\cite{Grimvall1981-s}
\begin{equation}\label{eqS1}
\widetilde{\omega}({\bf
q},\nu,T)=\texttt{Re}\left[\widetilde{\omega}({\bf
q},\nu,T)\right]-i\texttt{Im}\left[\widetilde{\omega}({\bf
q},\nu,T)\right],
\end{equation}
with the energy as the real part
\begin{equation}\label{eqS2}
\texttt{Re}\left[\widetilde{\omega}({\bf
q},\nu,T)\right]=\omega_{0}({\bf q},\nu)+\Delta\omega_{V}({\bf
q},\nu;T)+\Delta\omega_{pp}({\bf q},\nu;T),
\end{equation}
and the line width (half width at half maximum) as the imaginary
part
\begin{equation}\label{eqS3}
\texttt{Im}\left[\widetilde{\omega}({\bf
q},\nu,T)\right]=\gamma_{ep}({\bf q},\nu)+\gamma_{pp}(\bf{q},\nu;T),
\end{equation}
where $\omega_{0}(\bf{q},\nu)$ is the harmonic phonon energy at
$T=0$ K including the $T$-independent EPC contribution,
$\Delta\omega_{V}(\bf{q},\nu;T)$ is the energy shift due to the
change in the volume, and $\Delta\omega_{pp}(\bf{q},\nu;T)$ is the
energy shift due to the anharmonic phonon-phonon interactions. The
imaginary term $-i\texttt{Im}\left[\widetilde{\omega}({\bf
q},\nu,T)\right]$ describes the damping of the phonons, with
contributions from both EPC $\gamma_{ep}(\bf{q},\nu;T)$ and
anharmonicity $\gamma_{pp}(\bf{q},\nu;T)$.

To obtain the EPC constant from Allen's formula, the EPC-induced
line width $\gamma_{ep}$ is a prerequisite quantity. Thus extracting
the contribution of EPC from measured phonon line widths is
necessary. Up to now, however, it is still a challenge in practice
to directly obtain the pure EPC induced phonon line widths from
experiments. This is mainly because the measured phonon line widths
will probably also contain additional contributions from anharmonic
phonon-phonon interaction. Only when the temperature-dependent
anharmonic phonon-phonon interaction is correctly deducted, can the
EPC-induced line width $\gamma_{ep}$ be determined and $\lambda$ be
calculated.

Here we consider three different phonon decay channels to model the
anharmonic effect:

(1) An optical phonon with energy $\hbar \omega_0$ decays into two
acoustic phonons with energy $\hbar \omega_0/2$, which is a three
phonon decay process \cite{Hart1970-s}. In this case the
temperature-dependent phonon energy and line width can be expressed
as:
\begin{equation}
\hbar\omega(T)=\hbar \omega_{0}+\hbar \omega_{a}(1+\frac{2}{e^{\hbar
\omega_0/2k_BT}-1})
\end{equation}
\begin{equation}
\gamma(T)=\gamma_{ep}(1+\frac{2}{e^{\hbar \omega_0/2k_BT}-1})
\end{equation}

(2) An optical phonon mode with energy $\hbar \omega_0$ decays into
three acoustic phonon modes with energy $\hbar \omega_0/3$, which is
a four phonon decay process \cite{Balkanski1983-s}. In this case the
temperature-dependent phonon energy and line width can be expressed
as:
\begin{equation}
\hbar\omega(T)=\hbar \omega_{0}+\hbar \omega_{a}(1+\frac{2}{e^{\hbar
\omega_0/2k_BT}-1})+\hbar \omega_{b}[1+\frac{3}{e^{\hbar
\omega_0/3k_BT}-1}+\frac{3}{(e^{\hbar \omega_0/3k_BT}-1)^2}]
\end{equation}
\begin{equation}
\gamma(T)=\gamma_{a}(1+\frac{2}{e^{\hbar
\omega_0/2k_BT}-1})+\gamma_{b}[1+\frac{3}{e^{\hbar
\omega_0/3k_BT}-1}+\frac{3}{(e^{\hbar \omega_0/3k_BT}-1)^2}]
\end{equation}

(3) An optical phonon mode with energy $\hbar \omega_0$ decays into
another two optical phonon modes with lower eneryies $\hbar
\omega_1$ and $\hbar \omega_1$ \cite{Menendez1984-s}. In this case
the temperature-dependent phonon energy and line width can be
expressed as:
\begin{equation}
\hbar\omega(T)=\hbar \omega_{0}+\hbar \omega_{a}(1+\frac{1}{e^{\hbar
\omega_1/k_BT}-1}+\frac{1}{e^{\hbar \omega_2/k_BT}-1})
\end{equation}
\begin{equation}
\gamma(T)=\gamma_{ep}(1+\frac{1}{e^{\hbar
\omega_1/k_BT}-1}+\frac{1}{e^{\hbar \omega_2/k_BT}-1})
\end{equation}

\begin{figure}[h!]
\begin{center}
\includegraphics[width=0.9\textwidth]{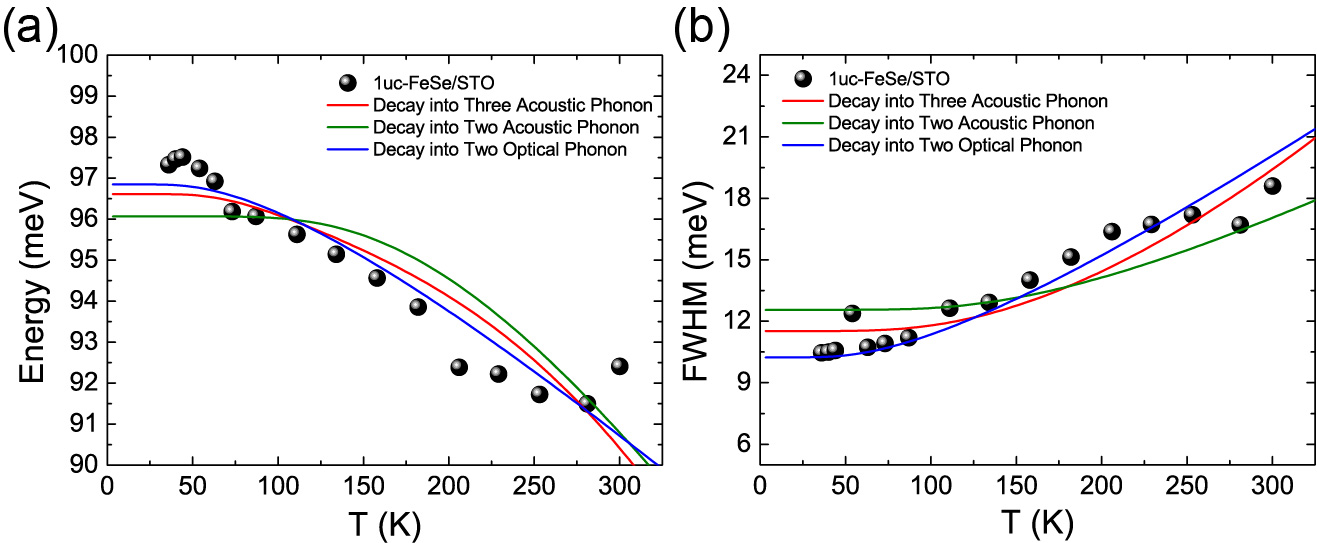}
\caption{\label{figS3} {\small The temperature dependence of the
energy (a) and full width at half maximum (FWHM) (b) of the $\alpha$
phonon branch of the 1uc-FeSe/STO sample. The dots are experimental
data obtained by fitting the EDCs from the HREELS spectra at the
$\Gamma$ point with Gaussian peaks, and the solid lines stands for
anharmonic decay fits of the data (red: $\alpha$ phonon branch decay
into three acoustic phonon; green: $\alpha$ phonon branch decay into
two acoustic phonon; blue: $\alpha$ phonon branch decay into two
optical phonon). The line width in (b) was obtained by deconvoluting
the measured phonon line width with the elastic peak width, to
remove the contributions from the instrumentation broadening and
surface roughness.}}
\end{center}
\end{figure}

The fitting result of these three decay models for the $\alpha$ mode
are exhibited in Fig. \ref{figS3} (a) and (b). The third model, an
optical phonon (92 meV) decays into two optical phonons (60 and 32
meV phonon branch), provides best fitting results for $\alpha$ mode.

Therefore we can extract $\gamma_{ep}$ for $\alpha$ mode through
above anharmonic phonon decay model. At $T=0$, there is no
phonon-phonon interaction, thus $\hbar\omega(T=0)$ and $\gamma(T=0)$
are the energy and line width with EPC only. As a result we have
extracted the EPC induced line width
$\gamma_{ep}(\alpha)=\gamma(T=0)\cong5.1$ meV. With $\gamma_{ep}$
and electron DOS, the EPC constant for $\alpha$ mode is deduced from
Allen's formula: $\lambda_{\alpha}\sim0.27$.

\subsection{Extracting EPC-induced phonon line widths from experimental results: An approach of Kramers-Kronig relation}

Some other indirect methods have also been tried to extract the
EPC-induced phonon line widths from experimental measurements. The
renormalization of the phonon energies and the line widths are
related by a Kramers-Kronig relation (or sometimes called Hilbert
transformation). If the unharmonicity is corrected (or showed to be
negligible), the EPC-induced line widths can be easily extracted.
Such an example is shown in a recent study of the electron-phonon
coupling on the surface of topological insulators
\cite{Zhu-PRL-2012-s}. Here we will employ a similar approach to
extract the EPC-induced line width in FeSe/STO system and compare
the results with anharmonic phonon decay method.

\begin{figure}[h!]
\begin{center}
\includegraphics[width=0.9\textwidth]{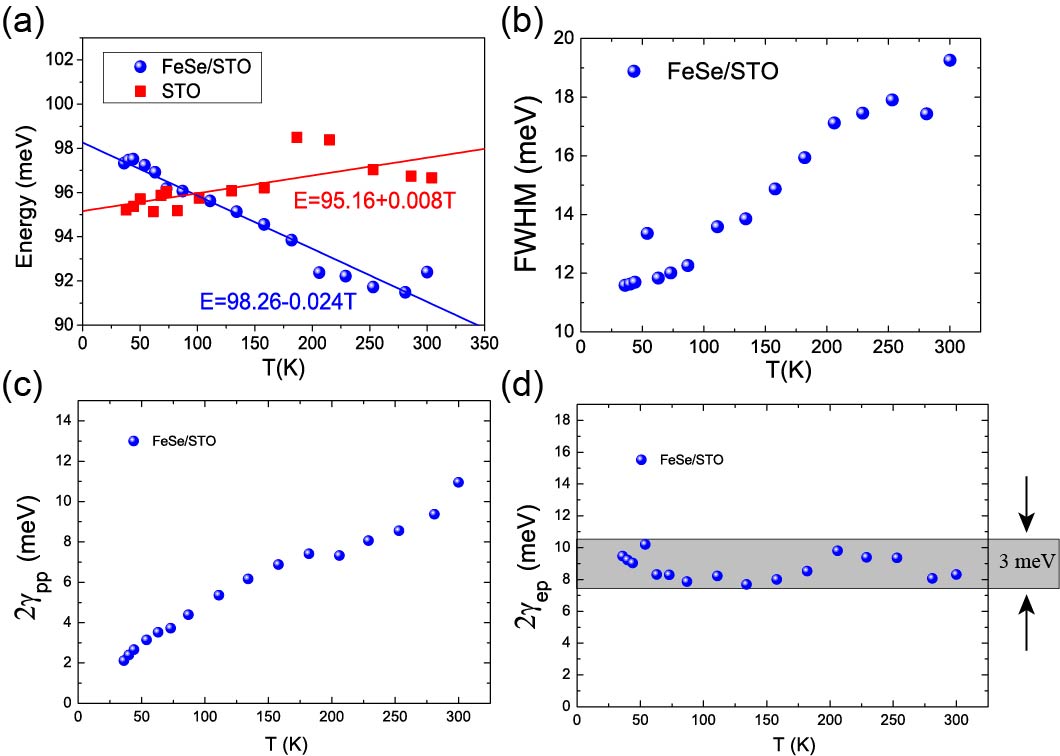}
\caption{\label{figS4} {\small The temperature dependence of the
energy (a) and full width at half maximum (FWHM) (b) of the $\alpha$
phonon mode, blue for 1uc-FeSe/STO, and red for bare STO. The dots
are experimental data obtained by fitting the EDCs from the HREELS
spectra at the $\Gamma$ point with Gaussian peaks, and the solid
lines stands for linear fits of the data. The line width in (b) was
obtained by deconvoluting the measured phonon line width with the
elastic peak width, to remove the contributions from the
instrumentation broadening and surface roughness. The line width
contribution from anharmonic phonon-phonon interactions (c) and from
EPC (d) of the $\alpha$ mode. The gray band in (d) represents the
energy resolution of our facility.}}
\end{center}
\end{figure}

Again we focus on the $\alpha$ phonon branch that almost has no
dispersion, then the phonon energy is nearly $\bf{q}$-independent.
Moreover, since the thickness of FeSe film is only 1 u.c. (about 0.5
nm), thus the energy shift due to the volume change
$\Delta\omega_{V}(\alpha;T)$ is mainly from the STO substrate. Here
we employ a simple linear fitting to both the STO and 1uc-FeSe/STO
data, as shown in Fig. \ref{figS4} (a), and obtain:
\begin{align}
&\Delta\omega_{V}(\alpha;T)+\Delta\omega_{pp,STO}(\alpha;T)=0.008T\\
\nonumber
&\Delta\omega_{V}(\alpha;T)+\Delta\omega_{pp,STO}(\alpha;T)+\Delta\omega_{pp,FeSe}(\alpha;T)=-0.024T
\nonumber
\end{align}
So the energy shift due to the unharmornic phonon-phonon
interactions from the FeSe film is:
\begin{equation}
\Delta\omega_{pp,FeSe}(\alpha;T)=-0.032T
\end{equation}

This energy shift can be rewritten in terms of energy (or frequency)
as:
\begin{equation}
\Delta\omega_{pp}(\alpha;T,\omega)=\frac{0.032T}{98.26-0.024T}\omega
\end{equation}

Then the Kramers-Kronig relation
\begin{equation}
\gamma_{pp}(\alpha;T,\omega)=\frac{2}{\pi}\int_0^{\infty}\frac{\omega\Delta\omega_{pp}(\alpha;T,\omega')}{\omega^2-\omega'^2}d\omega'
\end{equation}
can be used to calculate the line width contribution from unharmonic
phonon-phonon interactions. In real calculation, instead of using
infinity, we used the highest phonon energy $\omega_{max}=100$ meV
as a cutoff energy for the upper limit of the integral. The results
of $\gamma_{pp}$ are plotted in Fig \ref{figS4} (c). Then the
EPC-induced line width $\gamma_{ep}$ is obtained by subtracting
$\gamma_{pp}$ from the measured width in Fig. \ref{figS4} (b). As
expected, the resulting $\gamma_{ep}$ is very weakly
temperature-dependent, especially when our instrument resolution of
about 3 meV is taken into account. Consequently we obtain an average
EPC-induced line width $\overline{\gamma_{ep}(\alpha)}=4.5\pm0.5$
meV, yielding a EPC constant $\lambda_{\alpha}\sim0.24\pm0.03$.

So we have obtained similar $\gamma_{ep}$ as well as
$\lambda_{\alpha}$ from two independent methods, which indicates the
feasibility and reliability of our EPC constant calculation. Finally
we adopt an average value of the EPC constant for the $\alpha$ mode:
$\lambda_{\alpha}\sim0.25$.

\subsection{Extracting EPC-induced phonon line widths of $\beta$ branch}

\begin{figure}[h!]
\begin{center}
\includegraphics[width=1\textwidth]{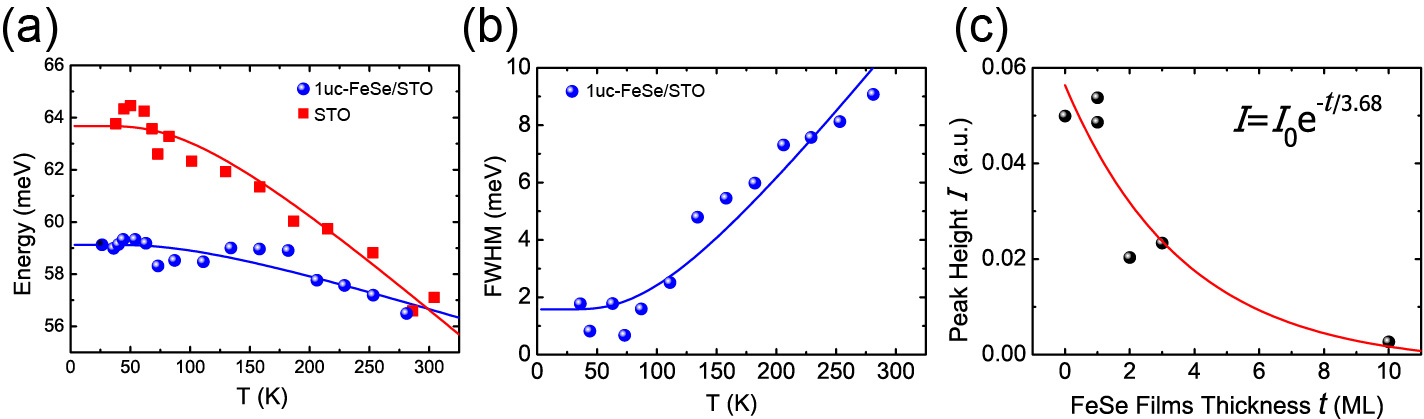}
\caption{\label{figS5} {\small The temperature dependence of the
energy (a) and full width at half maximum (FWHM) (b) of the $\beta$
phonon branch. The dots are experimental data obtained by fitting
the EDCs from the HREELS spectra at the $\Gamma$ point with Gaussian
peaks, and the solid lines stand for an anharmonic phonon fitting of
the data (red: STO; blue: 1uc-FeSe/STO). The line width in (b) was
obtained by deconvoluting the measured phonon line width with the
elastic peak width, to remove the contributions from the
instrumentation broadening and surface roughness. (c) Plot and
exponential fitting of the peak height of the $\beta$ mode as a
function of the FeSe thickness.}}
\end{center}
\end{figure}

The aforementioned approaches are also applied to the $\beta$ mode.
The same anharmonic phonon decay models are used to fit the
temperature-dependent energy and line width as well, with results
plotted in Fig. \ref{figS5} (a) and (b). Similarly, the third model,
an optical phonon ($\beta\sim60$ meV) decays into another two
optical phonons ($\zeta\sim32$ meV, and $\sigma\sim25$ meV at zone
boundary), provides the best fitting results.

The EPC-induced line width $\gamma_{ep}$ of $\beta$ mode is
extracted from Fig. \ref{figS5} (b): $\gamma_{ep}(\beta)\sim 0.8$
meV. And using Allen's formula, the corresponding EPC constant is
obtained: $\lambda_{\beta}\sim0.1$. This phonon mode had also been
observed as a replica band in ARPES spectra but can not be used to
calculate EPC constant due to the weak intensity
\cite{LeeJJ2014FeSereplica-s}. Whereas, HREELS provides a excellent
technique to detect surface elementary excitations and enables us to
acquire coupling constant between phonons from STO substrate and
electrons in FeSe films. The electric field of the $\beta$ mode can
also penetrate FeSe films. An exponential fit to the normalized peak
height of the $\beta$ mode, as a function of the FeSe thickness,
yields a decay length of $3.68\pm2.69$ u.c. (Fig. \ref{figS5} (c)),
which is similar to the penetration depth of the $\alpha$ mode. The
penetration of the $\beta$ mode actually provides one of the phonon
decay channels for the $\alpha$ mode in FeSe films.

Comparing with the $\alpha$ phonon mode, in which the growth of
single layer FeSe strengthens the anharmonic phonon decay, the
$\beta$ mode exhibits opposite tendency. The anharmonicity of the
$\beta$ mode of bare STO is stronger than that of 1uc-FeSe/STO,
since the decay of the $\beta$ mode energy as a function of
temperature of bare STO is stronger (shown in Fig. \ref{figS5} (a)).
Therefore, the anharmonic decay of the $\beta$ mode is much smaller
than that of the $\alpha$ mode, which is possibly an important
reason why $\lambda_{\beta}$ is smaller than $\lambda_{\alpha}$.

\end{document}